\documentclass[aps,prl,twocolumn]{revtex4-1} 
\usepackage{graphicx}    
\usepackage{amsmath}
\usepackage{xcolor}

\begin{document}

\title{Bursts in discontinuous Aeolian saltation}
\author{M. V. Carneiro$^1$}
\email{cmarcus@ethz.ch}
\author{K. R. Rasmussen$^2$  }
\email{geolkrr@geo.au.dk}
\author{H. J. Herrmann$^{1,3}$}
\email{hans@ifb.baug.ethz.ch}

\affiliation{$^1$ Computational Physics, IfB, ETH Z\"urich, Wolfgang-Pauli-Strasse 27, 8093 Z\"urich, Switzerland}
\affiliation{$^2$ Department of Geoscience, Aarhus University, Hoegh Guldbergsgade 2, 8000 Aarhus C, Denmark}
\affiliation{$^3$ Departamento de F\'isica, Universidade Federal do Cear\'a, 60451-970 Fortaleza, Cear\'a, Brazil}
\date{\today}

\begin{abstract}
Close to the onset of Aeolian particle transport through saltation we find in wind tunnel experiments a regime of discontinuous flux characterized by bursts of activity. 
Scaling laws are observed in the time delay between each burst and in the measurements of the wind fluctuations at the fluid threshold Shields number $\theta_c$.
The time delay between each burst decreases on average with the increase of the Shields number until sand flux becomes continuous. 
A numerical model for saltation including the wind-entrainment from the turbulent fluctuations can reproduce these observations and gives insight about their origin.
We present here also for the first time measurements showing that with feeding it becomes possible to sustain discontinuous flux 
even below the fluid threshold.

\end{abstract}
\maketitle


Sporadic bursts of sand are common wind erosion events. 
Through them, a granular surface can be slowly eroded even at rather low wind strength, contributing to long-term changes in the landscape \cite{Stout}. 
Bursts occur when the wind velocity fluctuates around the threshold controlling the initiation of sand movement.
Despite being quite common, little is known about this intermittent saltation mechanism.
 
In the laboratory, Bagnold observed that when the wind velocity above a quiescent sand surface increases above a certain limit some grains are for a short moment lifted into the air flow and gain momentum. 
When subsequently colliding with the bed and ejecting other sand grains saltation accelerates to an equilibrium value through a cascading reaction. 
The wind shear velocity of the wind flow at the minimum threshold for Aeolian particle entrainment Bagnold named the fluid threshold $u_t$ \cite{Bagnold1, Bagnold3}. 
Bagnold defined also a dynamic threshold. 
Once started, saltation will cease at a somewhat lower shear velocity due to momentum transfer from the impacting grains.
The ratio between the dynamic and the fluid threshold is 0.8

However, wind shear velocities above the fluid threshold can show temporarily no sand transport. 
Also Rasmussen and Sorensen observed saltation to be discontinuous below the fluid threshold in the field\cite{soresen_ras}. 
Intermittent fluctuations of the turbulent wind produce sand bursts, which can be observed individually if their density is low so that they do not superpose.
Induced by surface-layer turbulent structures, the sand bursts evolve downstream into coherent streamers \cite{Baas, Shao_Book}.
The burst effects of turbulence were first identified in sediment transport in water \cite{Heathershaw, Heathershaw2}. 
In aeolian saltation, Stout defined an intermittency function and found that bursts of saltation were approximately normally distributed with respect to the relative wind strength \cite{Stout}. 
Sch\"onfeldt statistically correlated the wind gusts in the velocity time series with the saltation events to propose varying dynamic and fluid thresholds.
All studies mentioned here were performed in the field, i.e. not under reproducible or fully controlled conditions. 
Numerical work of Dupont et. al. first to demonstrated a correlation between the flow turbulence and the sand streamers using Large Eddy Simulations (LES)\cite{dupont}.
The numerical observation of sand streamers requires large systems and, consequently several assumptions concerning particle interactions to facilitate the computational simulations. No previous works have provided yet an insight of sand bursts at the particle scale.

Here, we report on a systematic investigation of sand bursts not only experimentally observed in a wind tunnel but also numerically reproduced using a stochastic  process to model the temporal fluctuations in a turbulent channel flow.
The Discrete Element Model (DEM) is a flexible technique to include the main forces acting on a saltating particle and correctly reproduce the particle-particle and fluid-particle iterations. The first one is found in the particle splash and mid air collisions, and the second one is in the particle lift and particle drag \cite{Jump,midair}.

\begin{figure*} [ht]
{\includegraphics[scale=0.28]{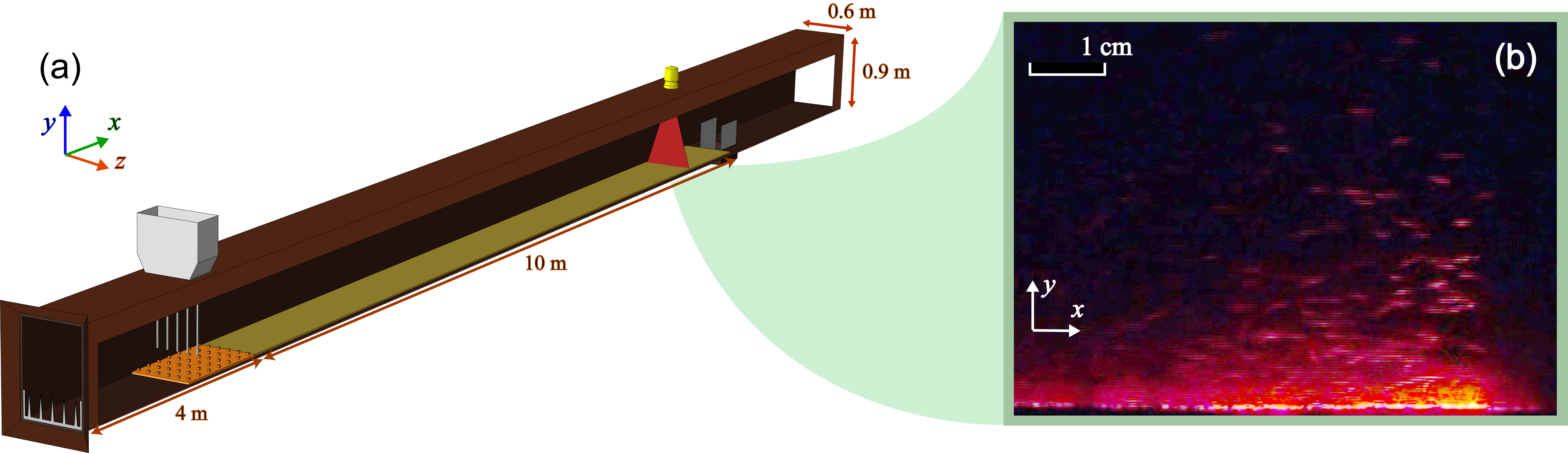}}
\caption{The experimental verification of sand bursts. (a) Experimental set-up of the wind tunnel. (b) A snapshot of the sand burst taken from the observation point at the downwind end of the wind tunnel. }
\label{tunnel}
\end{figure*}

At Aarhus university, we used a wind tunnel illustrated in Fig. \ref{tunnel}. 
The wind tunnel is 20 m long with 10 m of working section in the downstream end.
The rectangular cross section is $0.60$ m wide and $0.90$ m high.
A set of turbulence spires at the entry and a 3 m long array of roughness blocks downwind of the spires have been designed to produce 
a boundary layer which has approximately the same characteristics as that forming over an infinitely long sand bed near the fluid threshold \cite{Iversen}.
Downwind of the array, the bottom of the tunnel is covered by a 15 mm thick bed of pre-sieved sand samples of diameter $D_{mean} =  180 \mu$m.
The sand bed at low transport rate allows fairly long experiments without sand depletion upwind. 
A small container at the end of working section collects the transported sand from which we obtain the average flux.

Sand feeding in the wind tunnel has two main purposes. 
First, it guarantees that no depletion of sand occurs at the upwind end of the bed. 
Sand is fed into the tunnel via 5 tubes that distribute it fairly evenly across the bed.  
The feeding rate is approximately the same as that by which sand is transported downstream the tunnel.

Second, feeding is also useful for studying saltation when the shear stress is above the dynamic threshold as it simulates an infinite upwind fetch. 
The input of particle through feeding at the upwind end of the tunnel triggers saltation by adding momentum in the system.

Saltation is recorded on a standard video camera (2 megapixels) outside the tunnel.
An observation point, consisting of a video camera and a pitot tube for wind measurements, is located at the downwind  end of the sand bed before the sand collector, as indicated in Fig. \ref{tunnel}.
This focuses on the central part of the tunnel where saltating particles are illuminated when they cross a vertical laser sheet.
Color frames are extracted from the videos (see Fig. \ref{tunnel}), converted to gray-scale [0,1], and then into binary black and white images using a pixel threshold of 0.06 \cite{Otsu}.
Some particles may be lost during this selection.
Saltation activity is observed with the increase of white pixels in the time series.
From the video records, we measure the number and the length of the bursts.
The video frames do not give access to the individual dynamics of the grains.
The saturated flux is not quantified by the number of white pixels.
We identify the start and end of a burst using a calibrated threshold to discriminate the burst activity from the noise, which is when the number of white pixels from one frame to the  next changes by more than 30.
We measure a time delay between two sand bursts if sand transport stops for more than 3 seconds, which was arbitrary chosen short enough to include long bursts with short delays between bursts at high wind velocities.

Initially the duration of experiments varied from 2-3 minutes (short-runs), but this produced very poor statistics.
Therefore two series of new experiments were made where data were taken over three consecutive periods of 20 minutes (long-runs). The tunnel settings and the environmental conditions are described in the Supplemental Material \cite{SM}.

In the numerical simulations, a three dimensional wind channel of dimensions $(700\times50\times7.5) D^3_{mean}$ with periodic boundary conditions in the direction of the wind contains a poly disperse 3D quiescent packing composed by 12 layers of hard spheres with Gaussian distributed diameters with $D_{mean} = 200 \mu$m, size dispersion $\sigma_D = 0.15D_{mean}$, and density $\rho_s = 2650$ kg/m$^3$ that are subjected to a gravitational field in vertical direction ($y$-direction) and a logarithmic wind velocity profile $u(y)$ imposed in horizontal direction ($x$-direction),

\begin{equation}
u(y) = \frac{u_*}{\kappa}\ln\frac{y - h_0}{y_0}.
\label{wind_profile_eq}
\end{equation} 

\noindent In Eq. (\ref{wind_profile_eq}), $y_0 = D_{mean}/30 $ is the roughness of the bed, $h_0$ the bed height, $\kappa = 0.4$ the von K\'arm\'an constant, and $u_*$ the wind shear velocity. The air density is $\rho_a=1.174$ kg/m$^3$.
We express the wind velocity through the dimensionless Shields number, defined as 
\begin{equation}
\theta = \frac{u^2_*}{(\rho_s/\rho_a-1)gD_{mean}}.
\end{equation}

Similarly, we defined the fluid threshold Shields number $\theta_c =  u_t^2/(\rho_s/\rho_a-1)gD_{mean}$ as the minimum wind shear velocity for sand transport. The sand samples in the wind tunnel and the numerical spheres are within the typical range of fine wind blown particles. We normalize the Shields number in our results with $\theta_c$ as it strongly depends on the weather conditions \cite{McKenna}. The turbulent flow velocity $\mathbf u$ splits into the mean stream velocity $\mathbf u(y)$ and a stochastic part $\mathbf u_s$:

\begin{equation}
\mathbf u =  \mathbf u (y)  +  \mathbf u_s.
\end{equation}
\begin{figure} [ht]
{\includegraphics[scale=0.32]{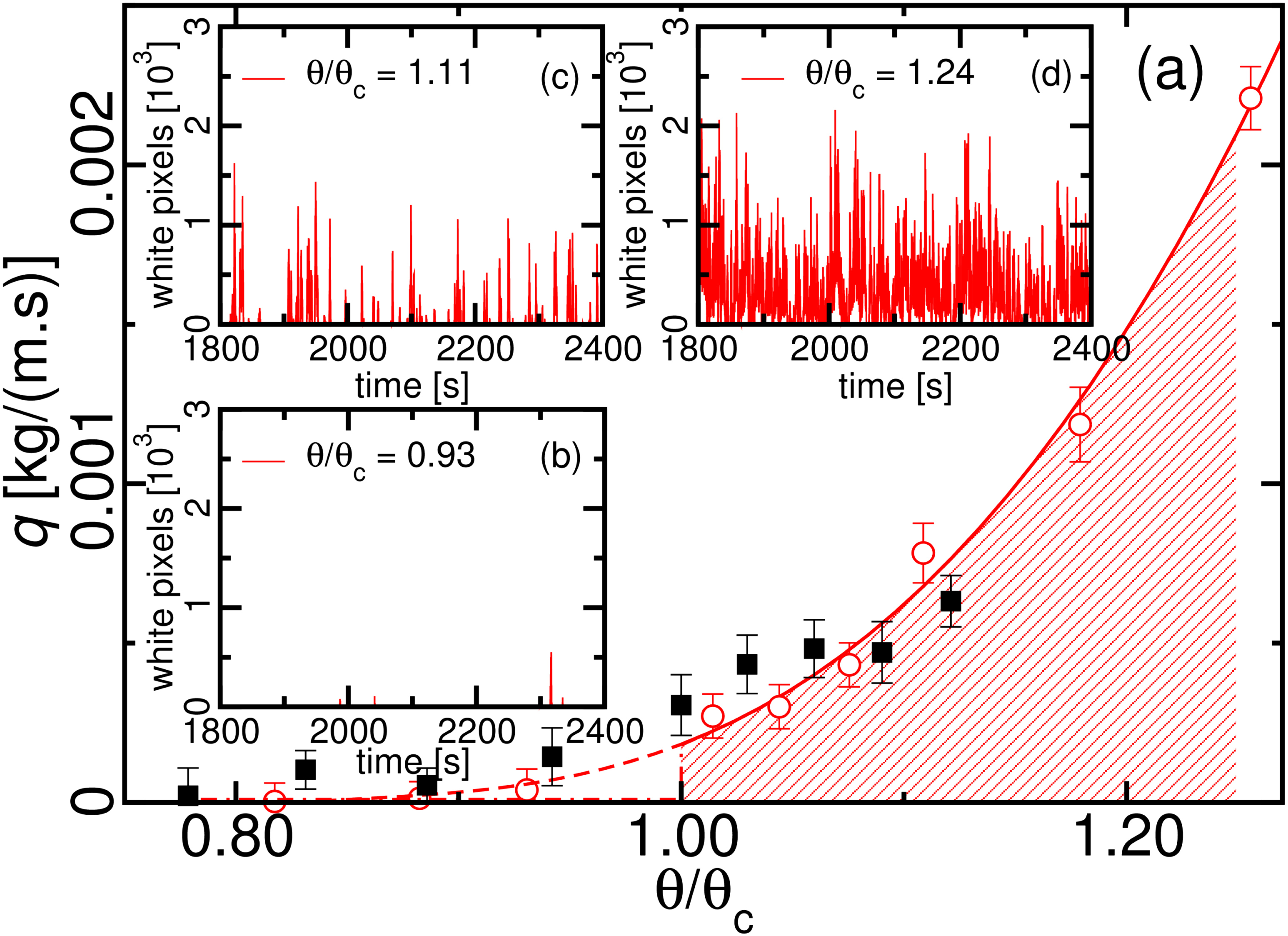}}
\caption{The sand transport rate in the long experiments. In (a), the transport rate obtained from the sand collected in the experiments (red circles) is compared with the numerical results  (black squares) using  $\langle\varepsilon\rangle = 0.05$ kg/(m.s$^3$).  Figures (b), (c) and (d) show the time series obtained from the video analysis.}
\label{long_runs}
\end{figure}

For $\mathbf u_s$, several measurements \cite {Porta, Mordant, Mordant2, Voth} have shown a highly non-Gaussian behavior of the Lagrangian acceleration distribution of fluid particles. 
Reynolds \cite{Reynolds} formulated a system of stochastic differential equations (SDE) for the logarithm of the dissipation rate, $\chi = log(\varepsilon/\langle\varepsilon\rangle)$, with $\langle\varepsilon\rangle$ as the mean dissipation rate that reproduces the Gaussian distribution of the velocities $u_s$ and the highly non-Gaussian distribution for the acceleration $a_s$ of the particles in fully developed turbulence.
Within reasonable bounds, the probability distribution of the flow perturbation has negligible effect on the results in steady state.
Every particle is attached to a tracer that generates random perturbations.
The dissipation rate $\langle\varepsilon\rangle$ controls the width of the probability distribution of these perturbations and consequently the rate of particle lift. Wider distributions produce stronger turbulent accelerations and lift more particles.
Details of the Discrete Element implementation, wind profile integration considering the momentum exchange between the wind and the grains, and the turbulence model are discussed in the Supplemental Material \cite{SM}.

The sand flux in the direction of the wind is defined as,

\begin{equation}
q  = \frac{1}{A}\displaystyle\sum\limits_{i}^N m_i v_{i}^{x},
\label{dimensionless_flux}
\end{equation}

\noindent where $A = (50 \times 7.5) D^2_{mean}$ is the area of the bottom of the channel, $v_{i}^{x}$ and $m_i$ are, respectively, the particle velocity in the $x$-direction and the mass of the particle $i$.
The saturated flux is the average flux in the stationary state.
The reinjection of particles at one side of the system domain after they cross the periodic boundaries might act effectively as a sand feed and increase the numerical predictions for the sand flux.


Without the sand feeding, the experiments in the wind tunnel for $\theta_c < 0.0072$ produced no grain  at the end of the wind tunnel.
The sand transport shoots up very steeply at $\theta_c \approx 0.0072 \pm 2 \times 10^{-5}$.
Figure \ref{long_runs}a shows the transported sand (red circles) collected at the container at the downwind end during the long runs. 
Within the error bars, saltation activity is very rare and the sand flux is virtually zero for $\theta/\theta_c = 0.93$ as shown in the flux series on Fig \ref{long_runs}b. 
The gradual increase of $\theta/\theta_c$ shows the build up of the continuous flux (Fig. \ref{long_runs}d) going through the discontinuous state (Fig. \ref{long_runs}c).
As the wind velocity increases, bursts superpose and the average burst size seen by the camera increases until the sand flux becomes continuous.
Therefore, the interval for a discontinuous saturated flux is $1 < \theta/\theta_c \lesssim 1.25 $. 
The numerical results from simulations (black squares) are discussed at the end of the manuscript.

\begin{figure} [ht] 
{\includegraphics[scale=0.32]{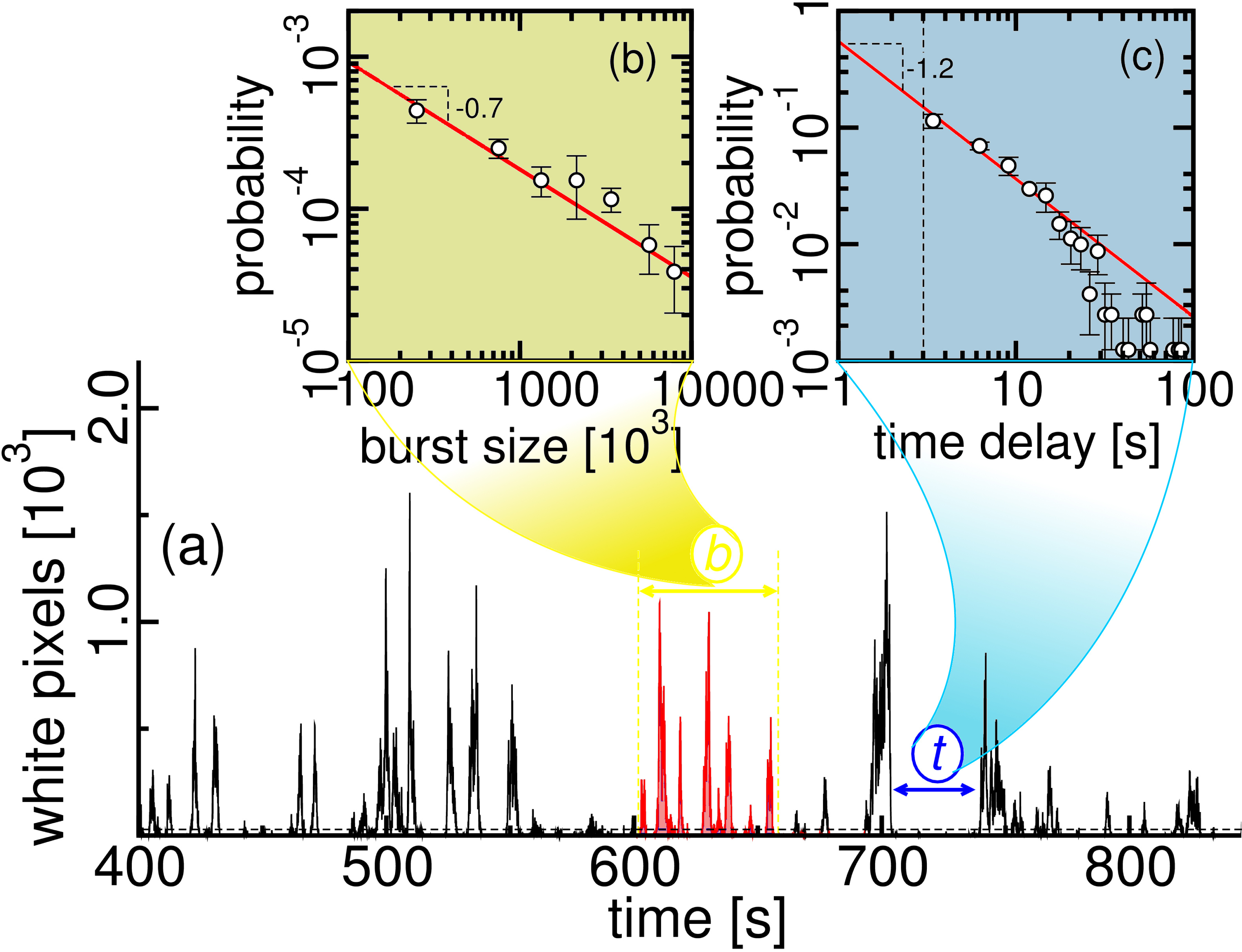}}
\caption{Scaling laws in discontinuous saltation. (a) The time series of intermittent saltation at $\theta_c$.
In (b), the burst size is obtained from the integration of the events within the yellow dashed lines in (a). The burst size distribution is fitted by a power law with an exponent $-0.7$. 
In (c), the time delays between the sand bursts illustrated by the blue arrow also follow a power law distribution with exponent $-1.2$.}
\label{bursts}
\end{figure}

Figure \ref{bursts}a shows the time series of the sand transport in the experiments at $\theta_c$.
An example of a sand burst is highlighted in red.
The video camera focuses on a narrow window of 75 mm length, which does not allow to observe multiple bursts.
The burst size in yellow and the time delay between bursts in blue display scaling laws.  
Figure \ref{bursts}b shows that the burst size distribution follows a power law with exponent $-0.7$.
The time delay (or waiting time) between bursts has also a power law distribution with exponent $-1.2$, as shown in Fig. \ref{bursts}c.
Scaling laws with comparable exponents can be also found in other complex systems with bursting activity, such as neuronal avalanches in rat cortex \cite{lombardi}.
At $\theta/\theta_c = 1.24$, the short and few intervals between bursts require much longer runs for getting good statistics.
The high frequencies of the distribution are bounded by the noise stemming from the video camera.
The low frequencies are bounded by the finite size of the wind tunnel which limits the formation of large low frequency eddies in the turbulent structure.  
Consequently, the probability distributions are restricted to a few orders of magnitude.
The bursting activity was not prominent enough to observe a power law for $\theta < \theta_c$.
Similar power laws fitted reasonably the time delays of the bursting activity for $\theta/\theta_c < 1.2$ \cite{SM}.
However, for $\theta/\theta_c > 1.2$, it is more difficult to obtain a clear distribution for the reducing time delays.

Whether a burst is generated locally from strong gusts at the area nearby or from the gradual development of the saltation initiated far upstream and propagating downstream cannot be determined unless we measure the velocity at a series of upstream observation points.
A sand burst observed in the videos can be the superposition of local bursts generated at different locations of the wind tunnel. However,  two different bursts do not have enough space to move alongside in the 0.6 m wide wind tunnel.
\begin{figure} [ht]
{\includegraphics[scale=0.34]{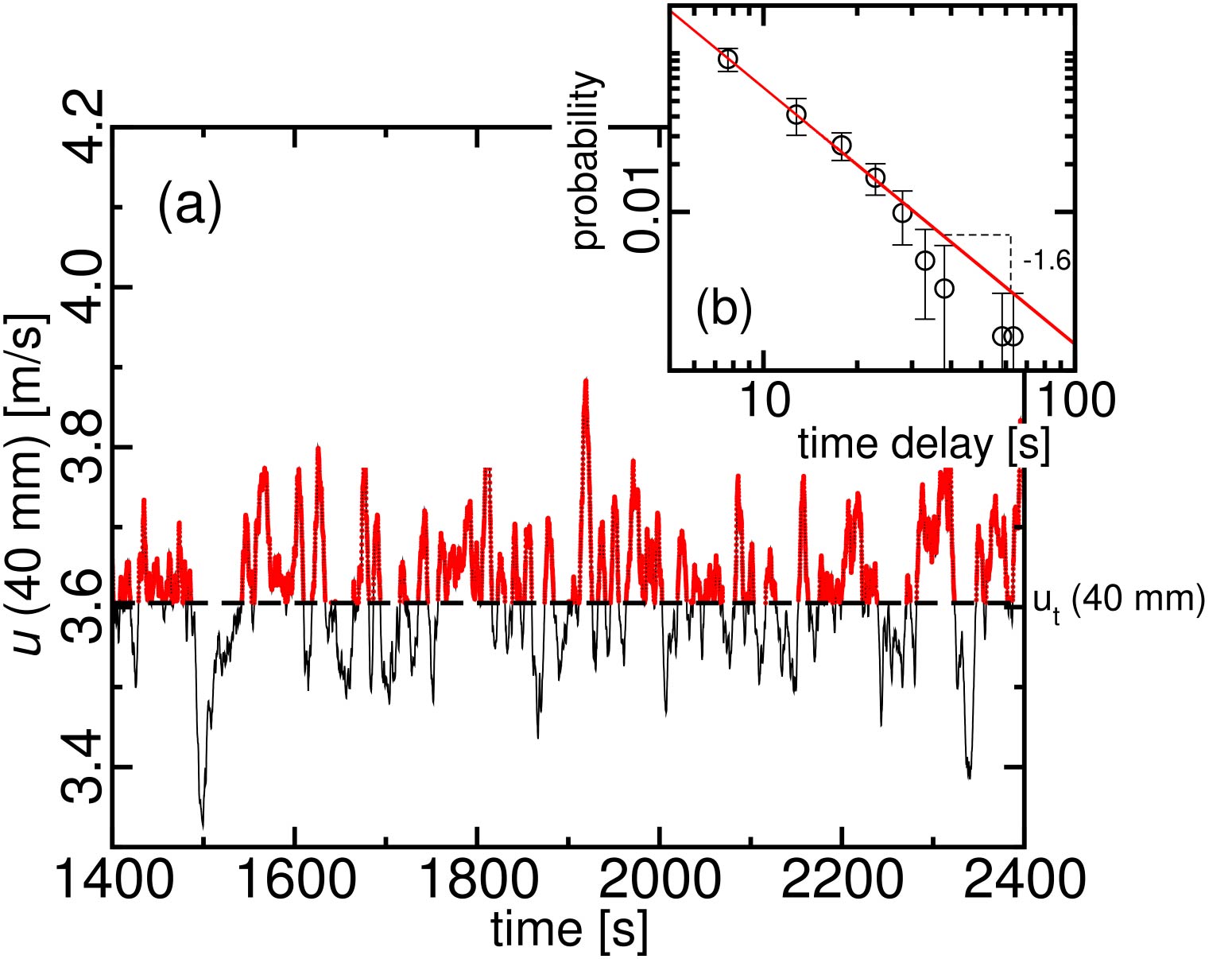}}
\caption{Scaling properties observed in the fluctuating wind at $\theta_c$. In (a), the wind fluctuations are measured at 40 mm above the sand bed. In (b), the probability distribution of the time delays between the wind fluctuations in red is fitted by a power law with exponent $-1.6$.}
\label{wind}
\end{figure}

The aerodynamic lift initiates the sand bursts. 
Figure \ref{wind} shows the wind velocity at the observation point for a run at $\theta_c$.
In red, we show the intervals where the wind velocity was larger than the average wind speed, i.e., the speed corresponding to $\theta_c$.  
The time delay/waiting time between wind fluctuations above the average velocity also follows a power law of exponent $-1.6$, as shown in Fig. \ref{wind}b.  
Similarities in the probability distributions suggest a correlation between the occurrence of the bursts and the fluctuations in the wind speed.
A definitive conclusion requires a complete data synchronisation between the wind measurements and the video records, which was unreachable with the state of art equipment and the collected material.
\begin{figure} [ht]
{\includegraphics[scale=0.34]{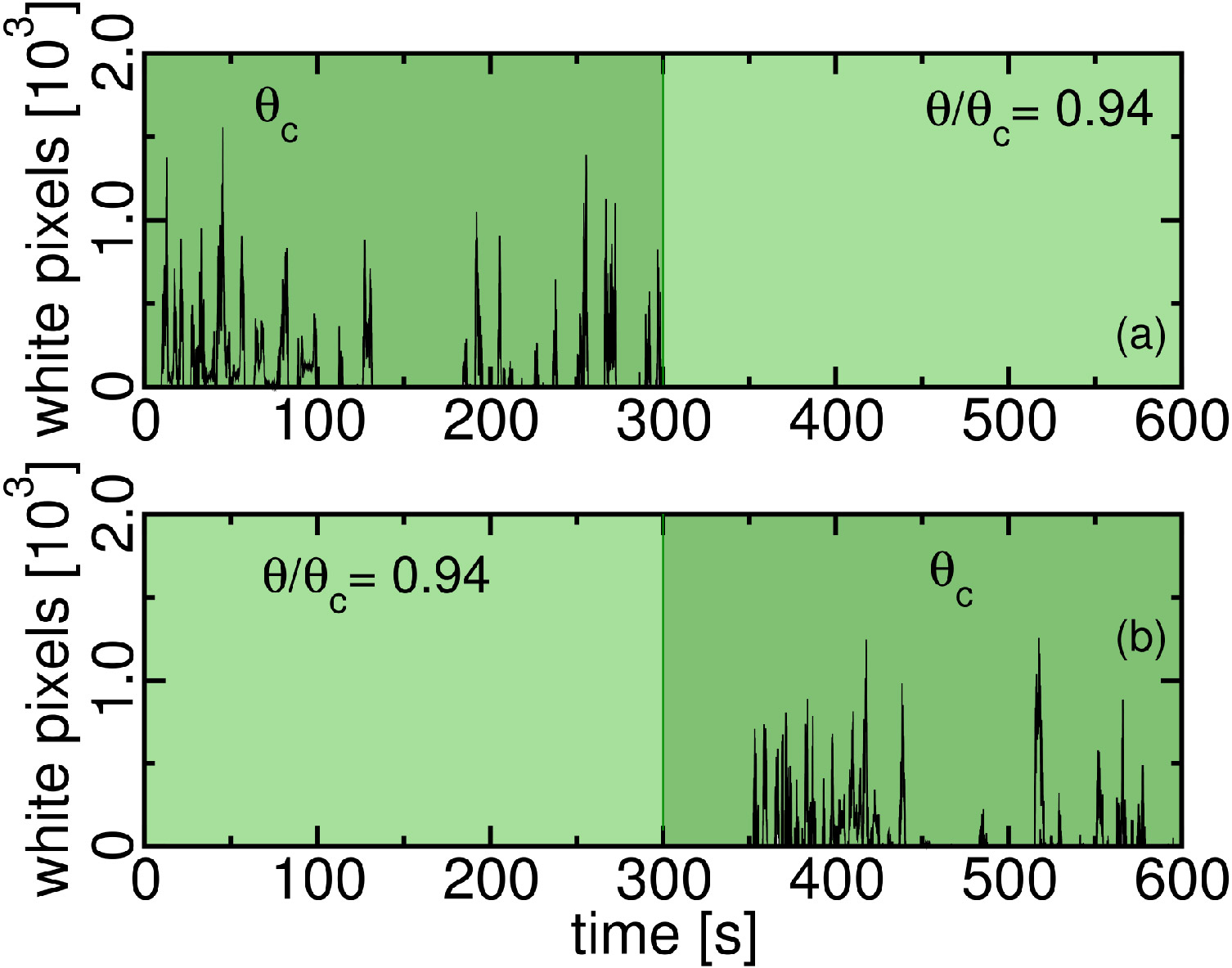}}
\caption{Sharp transition at discontinuous saltation at the neighbourhood of $\theta_c$ In (a), the bursting activity at $\theta_c$ stops when $\theta$ decreases to $\theta/\theta_c =0.94$. In (b), bursts start a few seconds after the Shields number increases to the fluid threshold Shields number $\theta_c$.}
\label{change}
\end{figure}

Figure \ref{change} gives insight about the triggering mechanism.
It shows, in $(a)$, that bursts disappear, when the Shields number decreases from $\theta_c$ to $\theta/\theta_c = 0.94$. Alternatively, the bursting appears after the Shields number increases from $\theta/\theta_c = 0.94$ to $\theta_c$, shown in $(b)$.

\begin{figure} [ht]
\includegraphics[scale=0.34]{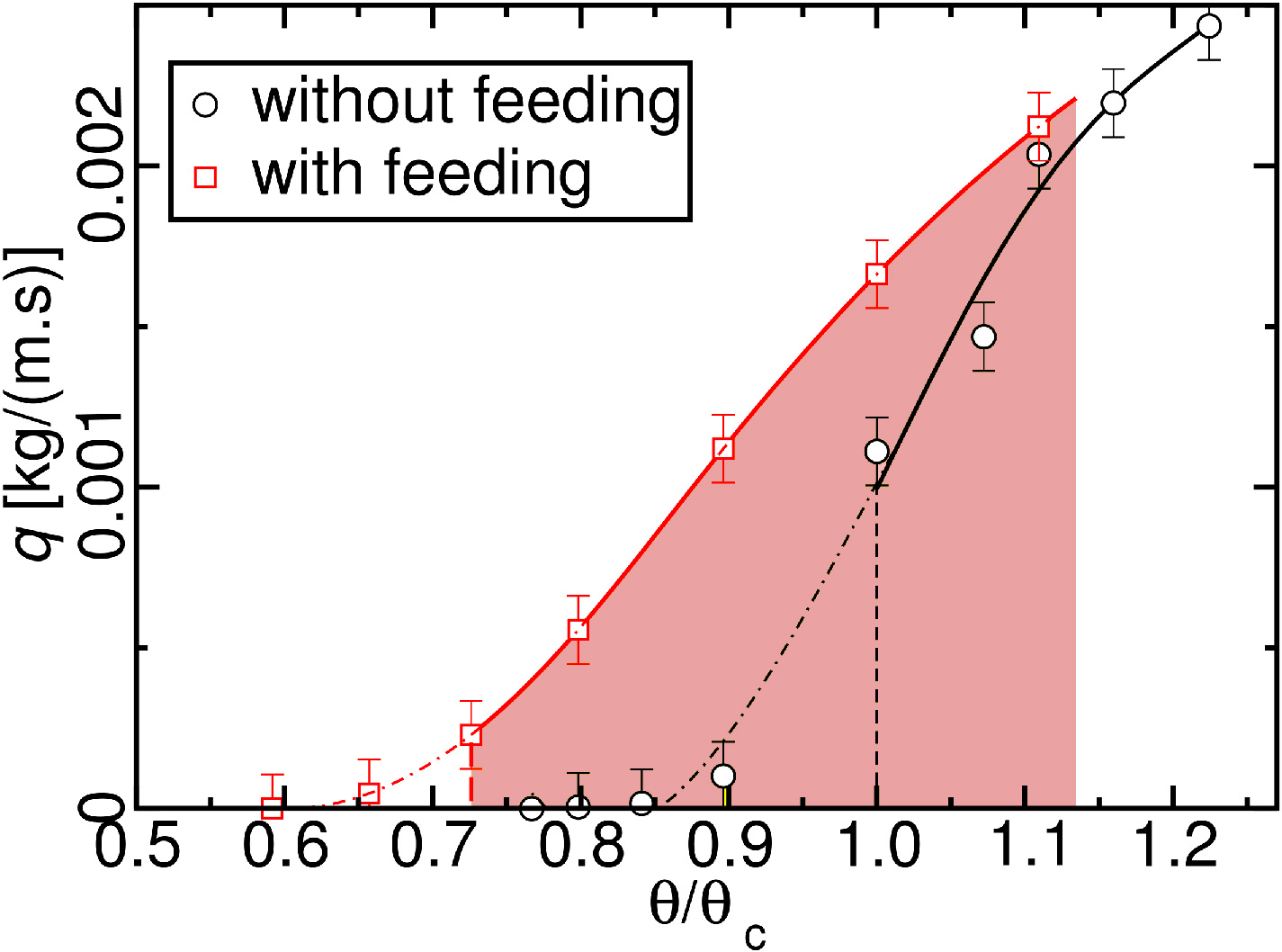}
\caption{The hysteresis zone between the fluid and dynamic thresholds. The short experiments with (red squares) and without feeding (black circles) show the hysteresis zone (red area) between  $\theta/\theta_c = 1.12$ and the dynamic threshold at $\theta/\theta_c = 0.73$.}
\label{hysteresis_exp2}
\end{figure}

Figure \ref{hysteresis_exp2} shows two different experiments, with (red squares) and without feeding (black circles).
With feeding it becomes possible to sustain a discontinuous flux even below the fluid threshold $\theta_c$. 
The occurrence of sand bursts is strongly dependent on the wind fluctuations around the fluid threshold. The momentum input from the sand feeding decreases the dependence of the sand bursts on the wind fluctuations and thus transport is sustained as long as the wind shear velocity is above the dynamic threshold.
Thus sand flux can occur for Shields numbers that did not display any transport without feeding.
The feeding significantly shifts the fluid threshold Shields number to $\theta/\theta_c = 0.73$.
The area in red shows the hysteresis zone that ends at  $\theta/\theta_c = 1.12$ where the transported sand is the same for both experiments within the error bars. 
The contribution of the feeding to the sand transport decreases as   $\theta/\theta_c \rightarrow 1.12$.
We note that the ratio between threshold Shields number with and without feeding is around $0.70$ (or $u_D/u_t = 0.84$).
\begin{figure} [ht] 
\includegraphics[scale=0.34]{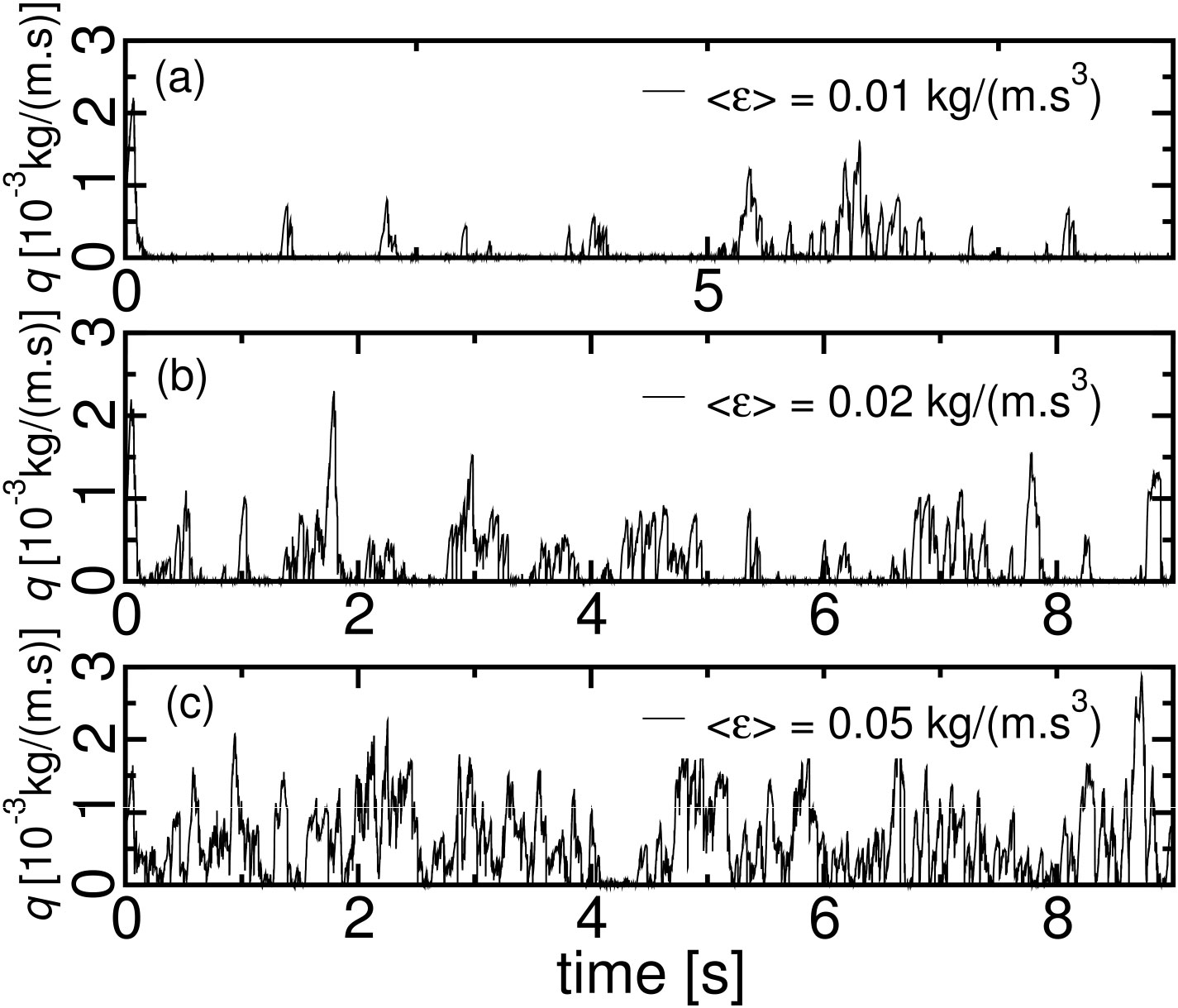}
\caption{Numerical simulation of intermittent saltation for $\theta/\theta_c = 1.02$ and different $\langle\varepsilon\rangle$. A completely settled particle bed, perturbations lift particles from the bed surface increasingly with $\langle\varepsilon\rangle$ and trigger saltation.}
\label{simulation}
\end{figure}

In the numerical simulations, the random turbulent perturbations mimic the wind pick of sand grains and reproduce the sand bursts.
Figure \ref{simulation}a shows the bursting activity at the sand flux at $\theta/\theta_c = 1.02$ and $\langle\varepsilon\rangle = 0.02$ kg/(m.s$^3$) at the numerical simulations.
From a completely settled particle bed, winds with dissipation rate $0.01$ kg/(m.s$^3$) $ < \langle\varepsilon\rangle < 0.06$ kg/(m.s$^3$) detach particles from the bed surface starting saltation.
If other particles are neither ejected nor lifted by the stochastic force meanwhile, saltation stops.
Saltation restarts if any other saltating particle is detached from the particle bed by the stochastic turbulent forces.
Increasing $\langle\varepsilon\rangle$, the lift rate increases and the time delay between sand burst decreases until sand flux becomes continuous.
Figure \ref{simulation}b shows the decrease of the time delays between bursts with the increase of $\langle\varepsilon\rangle$.
The flux is nearly continuous for $\langle\varepsilon\rangle = 0.05$ kg/(m.s$^3$), as shown by Fig. \ref{simulation}c.
Winds with dissipation rates $\langle\varepsilon\rangle  < 0.01$ kg/(m.s$^3$) entrain no particles, and $\langle\varepsilon\rangle > 0.06$ kg/(m.s$^3$) produce a continuous flux by entraining many particles.
Numerical simulations on the same time scale of the wind tunnel experiment are computationally too costly.
The qualitative results reproduce the phenomenology of the sand bursts on the particle scale.


In summary, the turbulent gusts create discontinuous saltation through bursting. A discontinuous saturated flux occurs at Shields numbers between $1 < \theta/\theta_c \lesssim  1.25 $. For $\theta/\theta_c \gtrsim 1.25 $, bursts superpose and the saturated flux becomes continuous.
Both wind measurements and video analysis displayed scaling laws and connect the burst initiation with the wind fluctuations above the fluid threshold.
The bursting of saltation could be numerically reproduced for the first time on a Discrete Elements Model (DEM) using stochastic process to model the temporal fluctuations of the wind turbulence.

We can not expect a perfect match between our findings in the wind tunnel and Nature. 
In fact, no wind tunnel is able to reproduce the low frequency of big eddies normally observed in the field. 
Therefore, the contribution of the bigger eddies can only be measured in field experiments. 
However, close to the onset of saltation bursts are still quite small and thus dominated typically by smaller turbulent eddies.

Additionally,  the same wind shear velocity could have different rates of mass transport in the field and in the wind tunnel. 
The wind shear velocity is a mean quantity which does not account the large temporal fluctuations of the wind.
The bursting of saltation, and consequently the transported sand, has a nonlinear dependence on wind fluctuations.

Aeolian sand bursts evolve downstream into streamers, which are so far, poorly understood. Future studies in the field should focus on these interesting coherent 3D structures.

We thank Nuno Ara\'ujo, Eric Parteli, Thomas P\"ahtz, Jens Jacob Iversen, Mohammad Hassani and Kornel Kovacs for discussions and acknowledge the Brazilian Council for Scientific and Technological Development (CNPq) and the European Research Council (ERC) Advanced Grant 319968-FlowCCS for financial support.

\section*{Author contributions}
M.C., K.R., and H.H. contributed equally to the work.

{\bf Competing financial interests:} The authors declare no competing financial interests



\end{document}